\documentclass[a4paper,12pt]{article}
\usepackage{graphicx}
\usepackage[T1,T2A]{fontenc}
\usepackage[utf8]{inputenc}
\usepackage{amssymb}
\usepackage{amsmath}
\usepackage{setspace}
\usepackage{xcolor}
\graphicspath{{images/}}
\begin{document} 

\thispagestyle{empty}

\begin{center}
{\large
{\bf 
Specific Features of Phase States of a Diluted 2D Magnet with Frustration
} 
}
\vskip0.5\baselineskip{
\bf 
D.~N.~Yasinskaya$^{*1}$, 
V.~A.~Ulitko$^{1}$,
Yu.~D.~Panov$^{1}$
}
\vskip0.1\baselineskip{
$^{1}$Ural Federal University 620002, 19 Mira Street,  Ekaterinburg, Russia
}
\vskip0.1\baselineskip{
$^{*}$daria.iasinskaia@urfu.ru
}
\end{center}

The properties of a dilute Ising magnet are studied using a two-dimensional spin–pseudospin model with charged impurities and a frustration caused by the competition of the charge and magnetic orderings. Based on the classical Monte Carlo method, the ground state phase diagram has been obtained and also unusual phase states appeared at finite temperatures have been studied. The regions in which order–order phase transitions and also reentrant phase transition are observed have been found.\\


\section{Introduction}
At present, the study of the properties of disordered and frustrated magnets and also diluted systems is of significant interest from both fundamental and practical points of view. Such systems have a rich ground state phase diagram and exhibit high sensitivity to external actions, demonstrating various types of phases and phase transitions at finite temperatures~\cite{Diep,Kaplan,Bramwell}. The interest in unusual phase states of such systems is particularly topical taking into account their close connection with spin liquids~\cite{Balents}.

The spin–pseudospin model considered in this work belongs to a wide class of the pseudospin Blume–Emery–Griffiths-type models~\cite{BEG} which are widely used for description of the properties of quantum and classical liquids, binary and ternary alloys, metamagnets, dilute magnets, cold atoms, and many other physical systems~\cite{BEG2}. The spin–pseudospin model was proposed in~\cite{Mod1} for the description of the competition between magnetic and charge orderings in HTSC cuprates in normal state. In this model for the cuprate CuO$_2$ plane along with [CuO$_4$]$^{6-}$ centers with spin 1/2, we consider the interacting [CuO$_4$]$^{5-}$
and [CuO$_4$]$^{7-}$ centers with spin 0 in the ground state. The more detailed justification of the model and the possibility of its using for describing the physical properties of cuprates and also the results obtained in the mean-field approximation are presented in~\cite{Mod2,Mod3}. The model Hamiltonian includes the on-site ($\Delta$) and intersite ($V$) density–density correlations for nonmagnetic
centers which have the form of single-ion anisotropy and the Ising exchange coupling in the language of pseudospin operators, and also the conventional spin
exchange interaction in the Ising form ($J$):
\begin{equation}
\mathcal {H} = 
\Delta \sum_i^{\phantom{N}} S_{iz}^2 
+ V \sum_{\left\langle ij\right\rangle} S_{iz} S_{jz} 
+ \tilde{J} \sum_{\left\langle ij\right\rangle}  \sigma_{iz} \sigma_{jz} - \mu \sum_i^{\phantom{N}} S_{iz}.
\label{H}
\end{equation}
Here, $\sigma_{iz} = (1-S^2_{iz}) s_{iz}$, $\tilde{J} = Js^2$, and $\mu$  is the chemical potential that is necessary to take into account the constant charge constrain
\begin{equation}\label{constrain}
n = \frac{1}{N} \sum_i \hat{S}_{iz} = const,
\end{equation}
where $n$ is the charge density. 
The states corresponding
to two pseudospin projections $S_z = \pm 1$ belong to two nonmagnetic [CuO$_4$]$^{5-,7-}$ states with charges $\pm 1$, which are counted from the charge of the magnetic [CuO$_4$]$^{6-}$  states with  $S_z=0$. The magnetic state with $S_z = 0$  is the spin doublet $s=1/2$. The summation is performed over the sites of a two-dimensional square lattice, ${\left\langle ij\right\rangle}$ denotes the nearest neighbors).

The term ``frustration'' can have various meanings; thus, it should be refined in this work. The systems with nonzero entropy of the ground state can be called frustrated systems~\cite{Kassan}. In the system considered in this work, nonzero entropy of the ground state is observed over a wide range of parameters due to the
existence of strongly interacting impurities~\cite{Shadrin}. In this work, we call the frustration point the values of the model parameters at which the ordering type is changed from the charge ordering to the magnetic ordering in the ground state. It is the classical analog of the quantum critical point.

A feature of our model is the existence of both the disorder (annealed charged impurities) and also the frustration related to the competition of various-type
interactions. In~\cite{Dotsenko,Giacomin}, it was shown that the introduction of impurities and various structural defects in the system substantially influences the phase states and the critical behavior, and also extends the possibilities of applying this model for the description of real physical systems. Thus, it is interesting to study the influence of charged impurities on the phase states of
our system near the frustration point.

\section{Methods}
The numerical simulation was performed using the classical Monte Carlo (MC) method. The charge constraint (\ref{constrain}) is provided by the modification of the
Metropolis algorithm~\cite{PAVT}. The calculations were carried out on a square lattice with periodic boundary
conditions, linear sizes La, and the number of sites
$N = L \times L$, where $a$ is the lattice constant taken to be 1.
All the calculations were carried out for the lattice $L=64$ with annealing of $1 \cdot 10^6$ MC steps per site, and the
data were averaged over 100 copies of the system. All
the effects and results discussed in this report were
checked for $L = 256$.

The temperature dependences of the specific heat
and the susceptibility are determined using fluctuation
relations
\begin{equation}
C=\frac{1}{N}\frac{\langle E^2 \rangle - \langle E \rangle^2}{k_BT^2}; \qquad
\chi = \frac{1}{N} \frac{\langle \mathcal{O}^2 \rangle - \langle \mathcal{O} \rangle^2}{k_B T}
\end{equation}
where $k_B$ is the Boltzmann constant and $E$ is the
energy of the system with Hamiltonian~(\ref{H}).
Order parameters $\mathcal{O}$ for the checkerboard antiferromagnetic and charge-ordered phases were determined as follows:
\begin{equation}
\mathcal{O} =
\left\{
\begin{array}{l}
m_1 - m_2,\\
M_1 - M_2.\\
\end{array}
\right.\\
\end{equation}
Here, $m_{\lambda}=\sum\limits_{i \in \lambda} s_{iz}$ is the magnetization of a sublattice $\lambda =1,2$, and $M_{\lambda}=\sum\limits_{i \in \lambda} S_{iz}$ -- is the summary charge of
a sublattice $\lambda$ (pseudo-magnetization).

The charge and spin structure factors were calculated using the following relations:
\begin{equation}
\begin{aligned}
&S(\vec{q}) = \frac{1}{N^2} \sum_{lm} e^{i\vec{q}(\vec{r}_l - \vec{r}_m)} \langle S_{lz} S_{mz} \rangle,\\
&s(\vec{q}) = \frac{1}{N^2} \sum_{lm} e^{i\vec{q}(\vec{r}_l - \vec{r}_m)} \langle s_{lz} s_{mz} \rangle,
\end{aligned}
\end{equation}

The critical phase transition temperatures were
determined by the maxima of the specific heat and the
susceptibilities. In this case, the error of determination
of the critical temperatures was no higher than 1
comparison with the critical temperatures found by
the Binder cumulant method. The existence of the
charge and antiferromagnetic orders (both long-range
and short-range orders) was determined using the values of the structure factors in point $a \vec{q} = (\pi,\pi)$.
\section{Results and discussion}
\begin{figure}[h!]
	\centering
	\includegraphics[width=\linewidth]{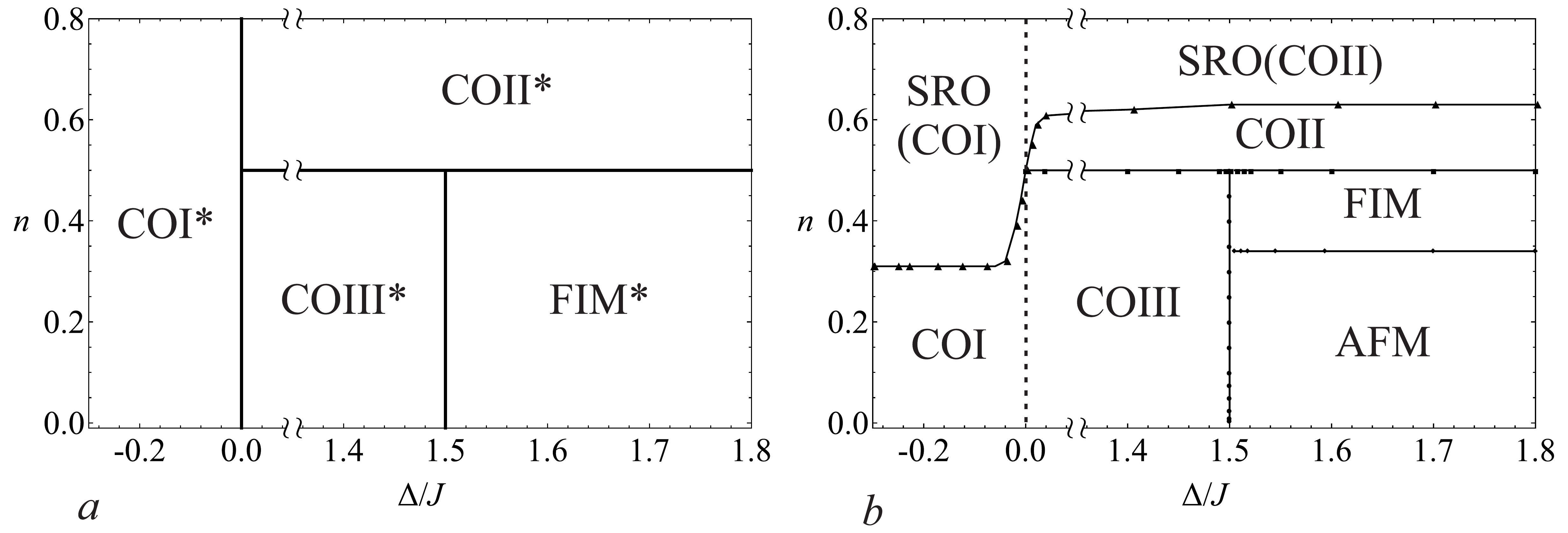}
	\caption{Ground state phase diagrams for the weak spin exchange limit obtained (a) in MFA and (b) using the MC method.}
	\label{GSPD}
\end{figure}

In~\cite{diffVJ}, it was shown that, in the mean-field
approximation (MFA), a change in the relation of
parameters $V$ and $J$ leads to two qualitatively different
ground state phase diagrams. In this work we have
restricted ourselves to the case of a weak spin exchange
($V = 4 \tilde{J}$). In this case, in MFA, four phases with the
checkerboard-type charge orderings (CO) form in the
ground state (Fig.~\ref{GSPD},a). COI phase corresponds to the
charge order without spin centers; COII and COIII
are the phases diluted with spin centers distributed
only over one sublattice. In COII phase at $\vert n \vert \geqslant 0.5$, one sublattice is completely filled with charge centers
of one type. In MFA, both phases have a magnetic
ordering that evidently is not observed during numerical MC calculations. MFA predicts the ferromagnetic
ordering for FIM phase; however, the numerical calculation showed the existence of a dilute antiferromagnetic (AFM) ordering at small $n$ and the shortrange AFM ordering at $0.33 \lesssim n < 0.5$. In addition, the
ground state phase diagram obtained in terms of the
MC method (Fig.~\ref{GSPD},b) differs in the existence of a large
region of the short-range charge ordering (SRO) of one or other type. The asterisks denote the phases
obtained in terms of MFA.

\begin{figure}[h!]
	\centering
	\includegraphics[width=\linewidth]{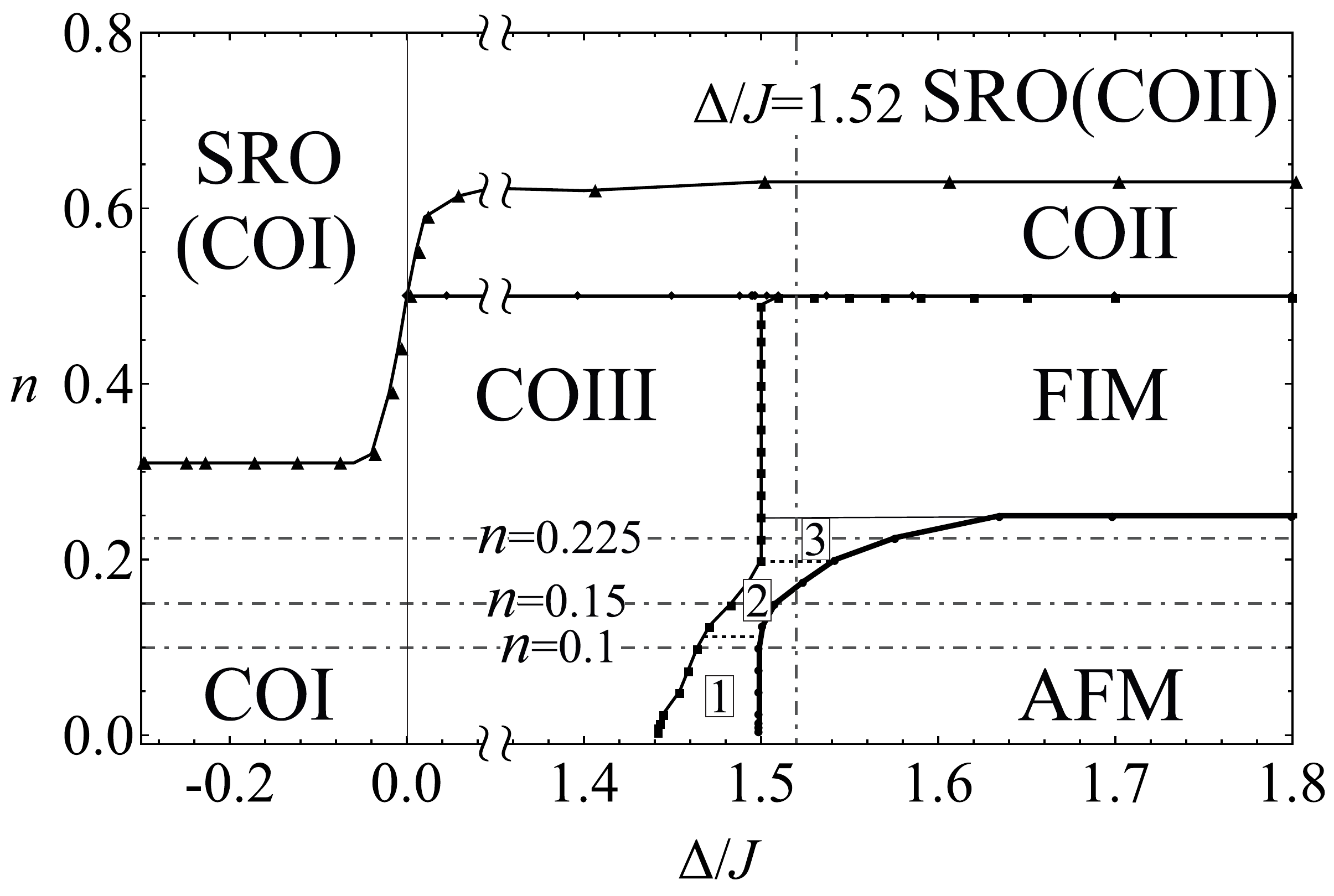}
	\caption{Diagram of possible phase states. Near the frustration point $\Delta^*/J=1.5$, there are regions 1, 2, and 3, in which the types of ordering are changed as the temperature decreases. The horizontal dash–dot lines correspond to the temperature phase diagrams shown in Figs.~\ref{n01}--\ref{n0225}. The vertical dash–dot line corresponds to the dependences of the structure factors on the charge density n in Fig.~\ref{SF}}
	\label{PD}
\end{figure}

Studying the phase states at finite temperatures is
of specific interest. The total diagram of possible
phase states for $V=4\tilde{J}$ is shown in Fig.~\ref{PD}. Near the
frustration point $\Delta^*/J=1.5$, there are three regions
labeled by numerals 1, 2, and 3 in frames in which the
ordering type is changed as the temperature decreases.

\begin{figure}[h!]
	\centering
	\includegraphics[width=\linewidth]{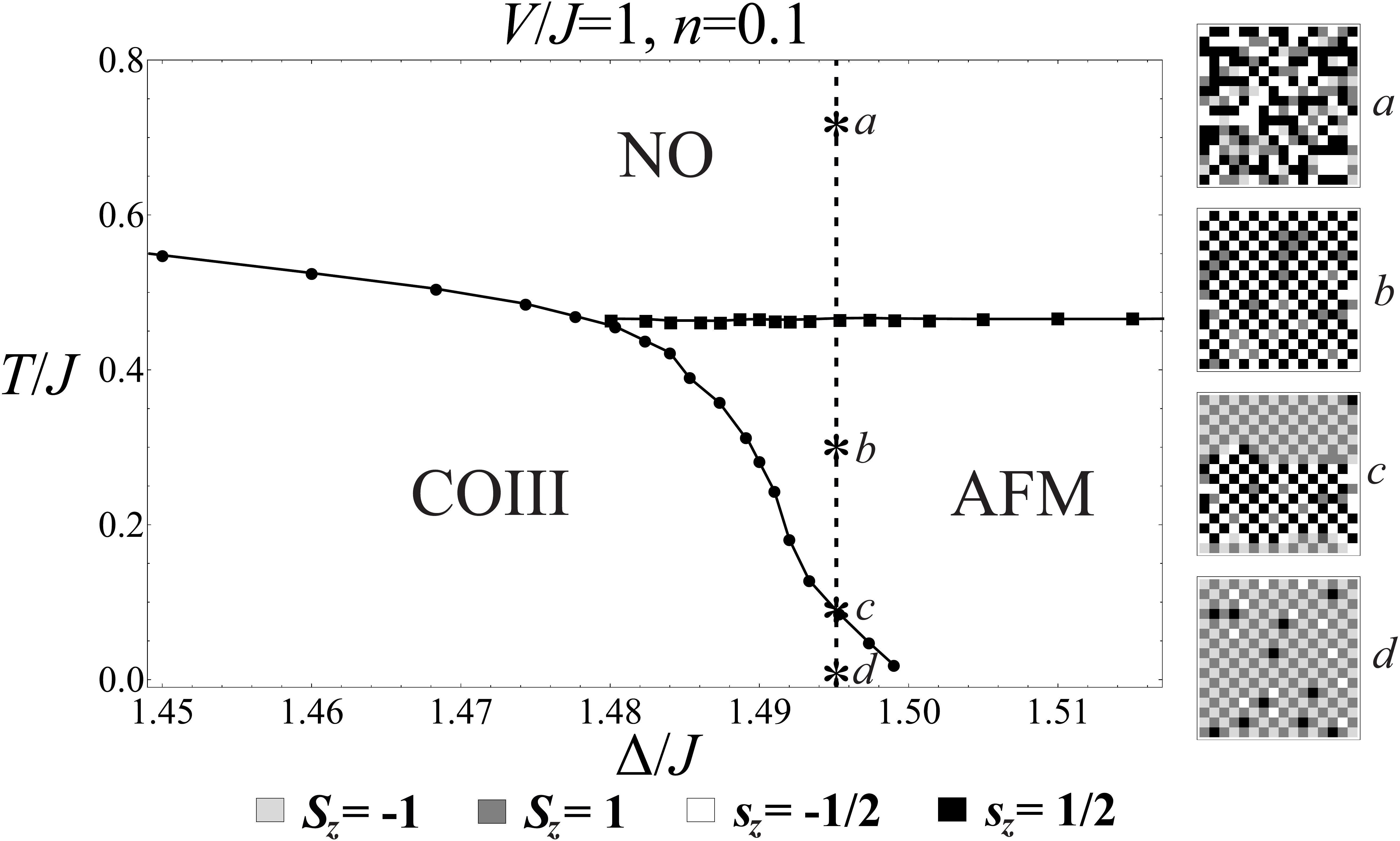}
	\caption{Temperature phase diagram for $n = 0.1$ near the frustration point. As temperature decreases, the second phase transition accompanied by changing the AFM ordering to the COIII phase, takes place.}
	\label{n01}
\end{figure}

\begin{figure}[h!]
	\centering
	\includegraphics[width=\linewidth]{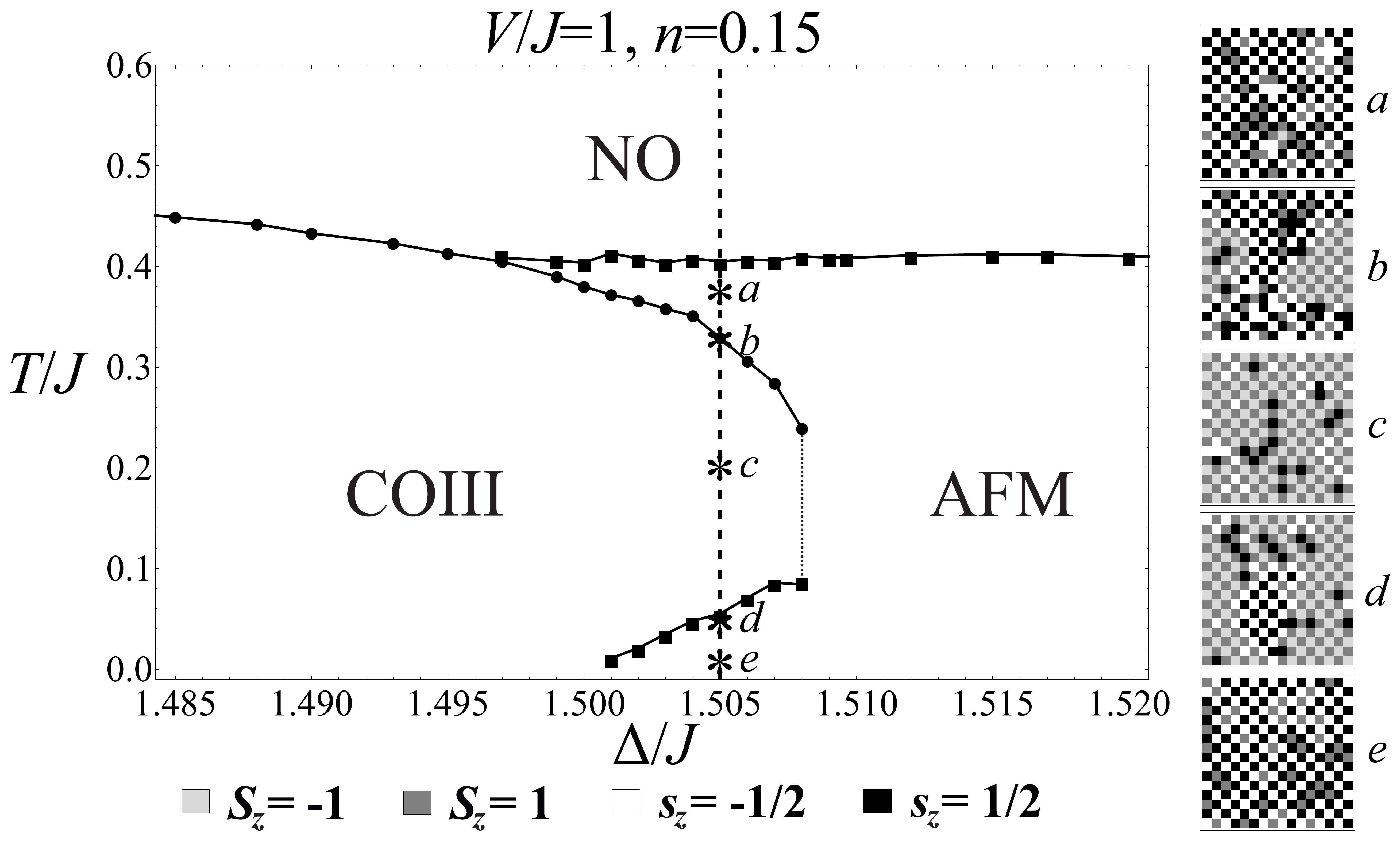}
	\caption{Temperature phase diagram for $n = 0.15$ near the frustration point. A double change in the ordering type is observed with an increase in temperature; i.e. the reentrant phase transition to AFM takes place.}
	\label{n015}
\end{figure}

\begin{figure}[h!]
	\centering
	\includegraphics[width=\linewidth]{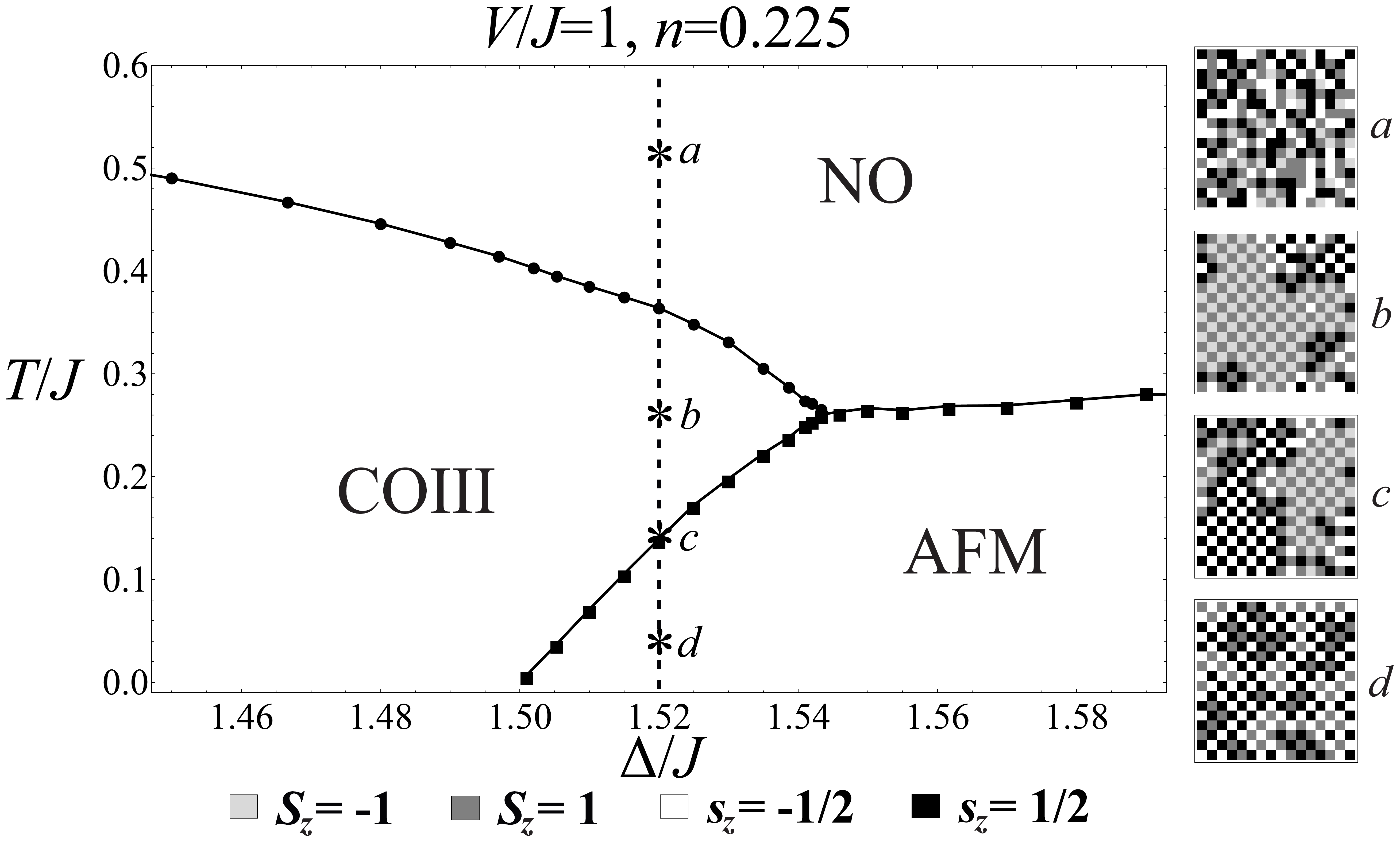}
	\caption{Temperature phase diagram for $n = 0.225$ near the frustration point. As temperature decreases, another phase transition from COIII to AFM is observed.}
	\label{n0225}
\end{figure}

Figures~\ref{n01}--\ref{n0225} show the temperature phase diagrams
for $n\,~{=}\,~0.1$, $n\,~{=}\,~0.15$, and $n\,~{=}\,~0.225$ corresponding to
the horizontal lines in regions 1, 2, and 3 in Fig.~\ref{PD},
respectively. At the right of the phase diagrams, there
are the snapshots of lattice fragments of $16 \times 16$ sites.
Different projections of pseudospin $S_z = \pm 1$ (two
charge states) and spin $s_z = \pm 1/2$ (two magnetic states)
are indicated by different gradations of the grey color.
In region 1 for $n=0.1$ (Fig.~\ref{n01}), a decrease in temperature leads to the phase transition from the nonordered state (NO) to the AFM phase and then, at
lower temperatures, another order–order phase transition with a change of the AFM ordering to the COIII
phase takes place. The change in the ordering type is
observed up to the value $\Delta/J=1.5$.

In region 2 (``bridge''), such a change in the orderings takes place twice as temperature decreases. This
situation is shown in Fig.~\ref{n015} for $n=0.15$. Thus, there are
three sequential phase transitions: from the high-temperature NO phase to the ordered AFM state; then,
the change to the COIII charge ordering, and the
reentrant phase transition to the AFM phase. In the
line for $n=0.225$ (Fig.~\ref{n0225}), we also observe the change
of one type of ordering to another, namely of COIII to
AFM at $\Delta/J>\Delta^*/J=1.5$.

\begin{figure}
	\centering
	\includegraphics[width=\linewidth]{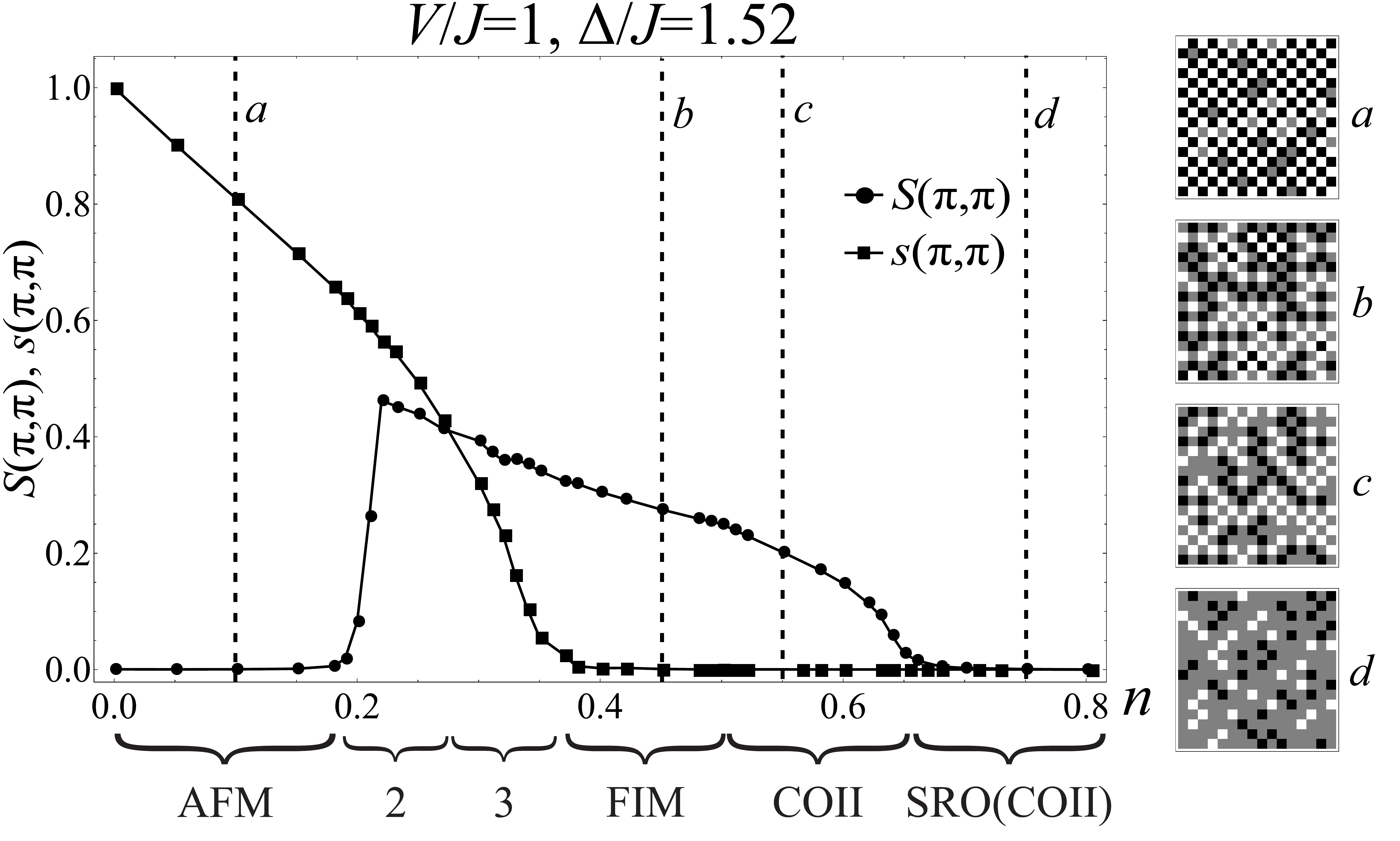}
	\caption{Dependence of the maxima of the charge and spin structure factors in point ($\pi,\pi$) on charge concentration $n$ for $\Delta/J = 1.52$.}
	\label{SF}
\end{figure}

Figure~\ref{SF} shows the dependence of the maximum
values of the charge and spin structure factors in point
($\pi,\pi$) on charge concentration $n$. This picture corresponds to vertical line $\Delta/J = 1.52$ in Fig.~\ref{PD}. The value
of the maximum of the spin (charge) structure factor
in point ($\pi,\pi$) allows us to judge the existence of the
AFM (CO) ordering and to determine the boundary
between the long-range and short-range orders. In this
case, we determined this boundary from the structure
factor value equal to 0.1. It is important to note that the
certain structure factor can reach its maximum value not in the ground state but at finite temperatures due
to the change in the ordering types. As shown in Fig.~\ref{SF},
at small $n$, a long-range order forms in the system. As
$n$ increases (regions 2 and 3), CO and AFM ordering
``coexist'' and change to one another. The condition of
conservation of the charge (\ref{constrain}) formally corresponds to
the existence of an external field acting to the pseudospin subsystem. In the case of a weak exchange at large
$n$, the charge ordering minimizes the energy of the
density–density correlations. In this sense, the charge
ordering is induced by the charge constraint. In the FIM phase, one sublattice is completely filled with
various magnetic centers and another sublattice is
filled with charge states of one type diluted with small
number of magnetic centers. Thus, the FIM phase
appears as the COII-type order diluted with the shortrange AFM order. It conserves up to $n=0.5$; in this
case, at $n \gtrsim 0.42$, the peak height of the spin structure
factor in point ($\pi,\pi$) is no higher than 0.002.

\section{Conclusions}
Using the classical MC method, we studied the two-dimensional spin–pseudospin model for the Ising magnet diluted with charged impurities and frustrated by the competition of the charge and magnetic orderings. The particular attention was focused on the influence of annealed charged impurities on the phase states of the system near to the frustration point in the case of a weak spin exchange.

It is shown that the competition between the charge and magnetic ordering leads to the formation of unusual phase states at finite temperatures. The MC-method simulation enabled us to refine the ground state phase diagram obtained in MFA before and to determine the boundary between the long-range and short-range orders. We also built the diagram of possible phase states that takes into account their evolution with an increase in temperature. Near the frustration point $\Delta^*/J=1.5$, we found three regions, in which the changes in the types of orderings and also reentrant phase transitions take place at finite temperatures. This effect is caused by the combination of the following factors. In the case of a weak spin exchange, the substantial concentration of charged impurities leads to the charge ordering, unlike the case of strong spin exchange~\cite{Mod2}, at which we observe the phase separation into macroscopic regions consisting of charge and magnetic centers. In addition, in the case of a weak exchange in the frustration point, the ground state is degenerate in energy for two various ordering types, namely, the charge and antiferromagnetic orderings. As a result, at finite temperatures near the frustration point, we can observe orderings that do not correspond to the minimum energy at $T=0$.
\section{Acknowledgement}
This work was supported by the program of enhancing the competitiveness of the Ural Federal University (Act 211 of the Government of the Russian Federation, Agreement no. $02.A03.21.0006$), the Ministry of Education and Science of the Russian Federation (project FEUZ-2020-0054, and also the Russian Foundation for Basic Research (project no. 18-32-00837$/18$).
%



\begin{thebibliography}{99}\setlength{\itemsep}{-0.10cm}
	
\bibitem{Diep} H. T. Diep, Frustrated Spin Systems, 2nd ed. (World Scientific, Singapore, 2013).

\bibitem{Kaplan} T. A. Kaplan and N. Menyuk, Philos. Mag. 87, 3711 (2007).

\bibitem{Bramwell} S. T. Bramwell and M. J. P. Gingras, Science (Washington, DC, U. S.) 294, 1595 (2001).

\bibitem{Balents} L. Balents, Nature (London, U.K.) 464, 199 (2010).

\bibitem{BEG} M. Blume, V. J. Emery, and R. B. Griffiths, Phys. Rev. A 4, 1071 (1971).

\bibitem{BEG2} V. V. Hovhannisyan, N. S. Ananikian, A. Campa, and S. Ruffo, Phys. Rev. E 96, 062103 (2017).

\bibitem{Mod1} Y. D. Panov, A. S. Moskvin, A. A. Chikov, and I. L. Avvakumov, J. Supercond. Nov. Magn. 29, 1077 (2016).

\bibitem{Mod2} Yu. D. Panov, V. A. Ulitko, K. S. Budrin, D. N. Yasinskaya, and A. A. Chikov, Phys. Solid State 61, 707 (2019).

\bibitem{Mod3} A. S. Moskvin and Yu. D. Panov, Phys. Solid State 61, 1553 (2019).

\bibitem{Kassan} A. V. Zarubin, F. A. Kassan-Ogly, and A. I. Proshkin, arXiv: 2002.05430 (2020).

\bibitem{Shadrin} A. V. Shadrin, V. A. Ulitko, and Y. D. Panov, J. Phys.: Conf. Ser. 1389, 012088 (2019).	

\bibitem{Dotsenko} V. S. Dotsenko, Phys. Usp. 38, 457 (1995).

\bibitem{Giacomin} G. Giacomin, in Proceedings of the \'{E}cole d'\'{E}t\'{e} de Probabilit\'{e}s de Saint-Flour XL2025, 2010.

\bibitem{PAVT} K. S. Budrin, V. A. Ulitko, A. A. Chikov, Yu. D. Panov, and A. S. Moskvin, in Proceedings of the Conference on Parallel Computational Technologies PCT'2018 (2018), p. 22.

\bibitem{diffVJ} Y. D. Panov, A. S. Moskvin, A. A. Chikov, and K. S. Budrin, J. Low Temp. Phys 187, 646 (2017).
\end{thebibliography}
\end{document}